\documentclass[preprint,aps,groupedaddress,nofootinbib,tightenlines]{revtex4}
\usepackage{graphicx,comment}
\usepackage{amsmath,amstext,amsfonts}
\begin{document}

\unitlength=1mm

\def\a{{\alpha}}
\def\b{{\beta}}
\def\D{{\Delta}}
\def\Dtilde{{\tilde{\Delta}}}
\def\d{{\delta}}
\def\e{{\epsilon}}
\def\g{{\gamma}}
\def\k{{\kappa}}
\def\l{{\lambda}}
\def\m{{\mu}}
\def\n{{\nu}}
\def\o{{\omega}}
\def\th{{\theta}}
\def\Dslash{D\hskip-0.65em /}
\def\vslash{v\hskip-0.50em /}
\def\Bslash{B\hskip-0.65em /}
\def\diag{\text{diag}}
\def\tr{\text{tr}}
\def\str{\text{str}}
\def\goesto{{\mathop{\longrightarrow}}}
\def\CPT{{$\chi$PT}}
\def\QCPT{{Q$\chi$PT}}
\def\PQCPT{{PQ$\chi$PT}}
\def\order{{\mathcal O}}
\def\Lag{{\mathcal L}}
\def\LQCD{{\Lambda_{\text{QCD}}}}

\preprint{NT@UW-02-029}

\title{Chiral $1/M^2$ 
       corrections to $B^{(*)}\rightarrow D^{(*)}$
       at Zero Recoil in \\ Quenched Chiral Perturbation Theory}
\author{Daniel Arndt}
\email[]{arndt@phys.washington.edu}
\affiliation{Department of Physics, Box 1560, University of Washington,
         Seattle, WA 98195-1560, USA}

\date{\today}

\begin{abstract}
Heavy quark effective theory can be used to calculate the values of
the semileptonic $B^{(*)}\to D^{(*)}$ decays
in the limit that the heavy quark masses are infinite. 
We calculate the lowest order chiral 
corrections, which are of ${\mathcal O}(1/M^2)$, from the breaking
of heavy quark symmetry 
at the zero recoil point in quenched chiral perturbation theory.
These results will aid in the extrapolation of 
quenched lattice calculations 
from the light quark masses used on the lattice
down to the physical ones.
\end{abstract}



\maketitle

\section{Introduction}
The Cabbibo-Kobayashi-Maskawa (CKM) matrix describes
the flavor mixing among the quarks, its elements are
fundamental input parameters for the standard model.
Their precise knowledge is not only  
crucial to determine the standard model
but also to shed light on the origin of $CP$ violation.
The matrix element that parameterizes the amount of mixing 
between the $b$ and $c$ quarks,
$V_{cb}$,
can be extracted from the
exclusive semileptonic $B$ meson decays 
$B\to Dl\nu$ and $B\to D^*l\nu$,
where $l=e,\mu$.
Heavy quark effective theory (HQET)
(for a recent review, see~\cite{Manohar:2000dt}),
which is exact in the limit of infinite masses $M$ for the heavy quarks,
predicts the width
of the process $B\to D^*l\nu$ as
\begin{equation}
  \frac{d\Gamma}{d\o}
  (B\to D^*)
  =
  \frac{G_F^2|V_{cb}|^2}{48\pi^3}
  {\mathcal K}(\o)
  {\mathcal F}_{B\to D^*}(\o)^2
,\end{equation}
where $\o=v\cdot v'$ is the scalar product of the 4-velocities 
$v$ and $v'$ of the
$D^*$ and $B$ mesons, respectively.
${\mathcal K}(\o)$ is a known kinematical factor and
${\mathcal F}(\o)$ is a form factor whose value at the kinematical
point $\o=1$ is ${\mathcal F}(1)=1$ in the $M\to\infty$ limit.
There are, however, perturbative and non-perturbative corrections 
to ${\mathcal F}(1)$,
\begin{equation}\label{eqn:F1}
  {\mathcal F}_{B\to D^*}(1)
  =
  \eta_A+\d_{1/M^2}+\dots
,\end{equation}
where the parameter $\eta_A\approx 0.96$ is a QCD radiative correction
known to two-loop order~\cite{Czarnecki:1996gu}
and $\d_{1/M^2}$ are non-perturbative corrections 
of ${\mathcal O}(1/M^2)$ to the
infinite mass limit of HQET.  
Note that, according to Luke's theorem~%
\cite{Luke:1990eg} there are no ${\mathcal O}(1/M)$ corrections
at zero-recoil.
One chooses the 
zero-recoil point 
because, for $\o=1$, ${\mathcal F}_{B\to D^*}$ can be expressed in
terms of a single form-factor $h_{A_1}$ given by
\begin{equation}
  \frac{\langle D^*(v,\e')|\bar{c}\gamma^\mu\gamma_5b|B(v)\rangle}
       {\sqrt{m_Bm_{D^*}}}
  =
  -2ih_{A_1}(1)\e'^{*\mu}
.\end{equation}
This is in contrast to the general case $\o>1$ for which
${\mathcal F}_{B\to D^*}(\o)$ is a linear combination of several
different form factors of $B\to D^*l\nu$ mediated by vector and
axial vector currents.

Several experiments,
most recently by CLEO~\cite{Briere:2002ew}, have determined the product
$({\mathcal F}_{B\to D^*}(1)|V_{cb}|)^2$ by 
measuring $d\Gamma_{B\to D^*}/d\o$ and extrapolating it to
the zero-recoil point.
The mixing parameter $|V_{cb}|$ can then be extracted once the value
${\mathcal F}_{B\to D^*}(1)$, that encodes the strong interaction physics,
has been evaluated.
The uncertainty in $|V_{cb}|$ is therefore determined by the 
experimental errors and by theoretical uncertainties in the
determination of ${\mathcal F}_{B\to D^*}(1)$.
Presently, the theoretical uncertainties dominate.%
\footnote{
Similarly, one can use the decay $B\to Dl\nu$ to extract
$({\mathcal F}_{B\to D}(1)|V_{cb}|)^2$ from the measured
$d\Gamma_{B\to D}/d\o$. However, $d\Gamma_{B\to D}/d\o$ is
more heavily suppressed by phase-space near $\o=1$ than 
$d\Gamma_{B\to D^*}/d\o$.
In addition, the $B\to D$ channel is experimentally more challenging.
Thus the extraction of $|V_{cb}|$ from this channel is less precise
but serves as a consistency check.
} 

A model-independent way
of calculating ${\mathcal F}(1)$
is provided by numerical lattice QCD simulations.
In this method one implements field theory
non-perturbatively using the Feynman path integral approach.
The fermion determinant that arises from the path integral
is very costly to calculate; it is often set to one in an
approximation called quenched QCD (QQCD).
This corresponds to dropping the contribution
from virtual quark loops
which are made of ``sea'' quarks
that have propagators not connected to 
the inserted external operators.
``Valence'' quarks, those that are connected to the inserted operators,
however, are kept.
Recently, such calculations have been performed%
~\cite{Simone:1999nv,El-Khadra:2001rv,Hashimoto:2001nb,Kronfeld:2002cc}
for the decays $B\to D^{(*)}l\nu$.
Several systematic uncertainties, 
such as from statistics and lattice space
dependence, contribute to the error of these calculations.
Another contribution to the uncertainties comes
from the chiral extrapolation
of the light quark mass.
Since lattice QCD simulations are limited by the available 
computing power they presently cannot 
be performed with the physical masses of
the light quarks. 
Therefore one needs to extrapolate from the heavier masses used
on the lattice (of order the strange quark mass)
down to the physical light quark masses. 
This extrapolation can be done by matching QQCD 
to quenched chiral perturbation theory (\QCPT)
and calculating the non-analytic corrections $\d_{1/M^2}$
in Eq.~(\ref{eqn:F1}) in \QCPT.
The formally dominant contributions to these corrections
come from the hyperfine mass splitting between the heavy pseudoscalar
and vector mesons that stems from the inclusion of
heavy quark symmetry breaking operators of
${\mathcal O}(1/M)$
in the Lagrangian.

In QCD, the corrections due to $D$~meson 
hyperfine splitting 
have been calculated in chiral perturbation theory (\CPT)
by Randall and Wise%
~\cite{Randall:1993qg}.  
A more complete treatment, 
involving additional corrections due to
$B$~meson hyperfine splitting,
${\mathcal O}(1/M)$ axial coupling corrections, 
and ${\mathcal O}(1/M)$ corrections to the current,
has been given in~\cite{Boyd:1995pq}.
Recently, the $D$ meson 
hyperfine splitting corrections have also been
determined in partially quenched
chiral perturbation theory (\PQCPT)%
~\cite{Savage:2001jw} 
for partially quenched QCD (PQQCD).
In contrast to QQCD,
PQQCD does not drop the contributions from sea quark loops
but gives different (and separately varied) 
masses to the sea and valence quarks.
Usually,
sea quarks are heavier than valence quarks, 
so that the fermion determinant --- although no longer equal to one ---
is much less costly to calculate than in ordinary QCD.

In this paper we calculate the ${\mathcal O}(1/M^2)$
corrections in Eq.~(\ref{eqn:F1})
due to $D$ and $B$~meson hyperfine splitting 
in quenched chiral perturbation theory (\QCPT).
As we will show,
these corrections are --- 
upon expanding in powers of the hyperfine splitting $\D$ ---
of order
$\LQCD^{3n/2}/(M^nm_q^{n/2})$ for $n\ge 2$
and formally larger than 
those coming from the inclusion of ${\mathcal O}(1/M)$
heavy quark symmetry breaking operators in the Lagrangian and
current which 
are suppressed by powers of 
$\LQCD/M$.
This argument is similar to the one that 
applies to \CPT~\cite{Manohar:2000dt}.

Our \QCPT\ calculation can be used to extrapolate lattice results%
~\cite{Hashimoto:2001nb} that
use the quenched approximation down to the physical light quark masses.
So far, this extrapolation has been based upon the \CPT\ calculation%
~\cite{Randall:1993qg}.
Using \QCPT\ 
should therefore give a better estimate of the uncertainties related 
to the chiral extrapolation.

A central role in the lattice calculation
of $B\to D^*$%
~\cite{Hashimoto:2001nb,Kronfeld:2002cc}
is played by the double-ratios of matrix elements
\begin{equation}\label{eqn:Rplus}
  {\mathcal R}_+
  =
  \frac{\langle D|\bar{c}\g^0b|B\rangle\langle B|\bar{b}\g^0c|D\rangle}
       {\langle D|\bar{c}\g^0c|D\rangle\langle B|\bar{b}\g^0b|B\rangle}
,\end{equation}
\begin{equation}\label{eqn:R1}
  {\mathcal R}_1
  =
  \frac{\langle D^*|\bar{c}\g^0b|B^*\rangle\langle B^*|\bar{b}\g^0c|D^*\rangle}
       {\langle D^*|\bar{c}\g^0c|D^*\rangle\langle B^*|\bar{b}\g^0b|B^*\rangle}
,\end{equation} 
and
\begin{equation}\label{eqn:RA1}
  {\mathcal R}_{A_1}
  =
  \frac{\langle D^*|\bar{c}\g^j\g_5b|B\rangle
        \langle B^*|\bar{b}\g^j\g_5c|D\rangle}
       {\langle D^*|\bar{c}\g^j\g_5c|D\rangle
        \langle B^*|\bar{b}\g^j\g_5b|B\rangle}
.\end{equation}
Since the numerator and denominator are so similar,
statistical fluctuations are highly correlated and cancel in the
ratios to a large degree.
The
${\mathcal O}(1/M^2)$ correction to the double ratios 
can therefore be calculated fairly
accurately and used to derive the 
${\mathcal O}(1/M^2)$ correction to the matrix elements
themselves.
For this reason, 
we also calculate ${\mathcal O}(1/M^2)$ corrections to the
decay $B^*\rightarrow D^*$
in addition to 
the  
experimentally accessible decays
$B\rightarrow D$ and $B\rightarrow D^*$,
and thus the corrections to 
${\mathcal R}_+$, ${\mathcal R}_1$, and ${\mathcal R}_{A_1}$.

\section{Quenched Chiral Perturbation Theory}
In a world where all the quark masses are large compared to
$\LQCD$ 
internal quark loops are suppressed and the results from 
QQCD are close to those from QCD.
In the real world, however, 
light quarks are light ($\ll\LQCD$) and contributions
from internal quark loops are substantial.
One can nevertheless study the low-energy behavior of QQCD
by its effective low energy theory, \QCPT%
~\cite{Morel:1987xk,Sharpe:1992ft,Bernard:1992mk,%
Bernard:1992ep,Golterman:1994mk}.
  
We consider a theory constructed from three light valence-quarks, 
$u$, $d$, $s$,
and three light bosonic quarks, 
$\tilde{u}$, $\tilde{d}$, $\tilde{s}$
governed by the Lagrangian of QQCD
\begin{equation}\label{eqn:LQQCD}
  {\cal L}
  =
  \sum_{a=u,d,s}\bar{q}^a(i\Dslash-m_q)^a_a q_a
  + \sum_{\tilde{a}=\tilde{u},\tilde{d},\tilde{s}}
      \bar{\tilde{q}}^{\tilde{a}}
      (i\Dslash-m_{\tilde{q}})^{\tilde{a}}_{\tilde{a}} 
      \tilde{q}_{\tilde{a}}
  =
  \sum_{j=u,d,s,\tilde{u},\tilde{d},\tilde{s}}
  \bar{Q}^j(i\Dslash-m_Q)^j_j Q_j
.\end{equation}
Here both types of quarks have been accommodated in the six-component
vector $Q$ with the three quarks $q_a$ in the upper three entries and the 
three bosonic ghost-quarks $\tilde{q}_{\tilde{a}}$ 
in the lower three entries.
The graded equal-time commutation relation governing the valence- 
and ghost-quarks is
\begin{equation}
  Q^\a_i({\bf x}){Q^\b_j}^\dagger({\bf y})
  -(-1)^{\eta_i \eta_j}{Q^\b_j}^\dagger({\bf y})Q^\a_i({\bf x})
  =
  \d^{\a\b}\d_{ij}\d^3({\bf x}-{\bf y})
,\end{equation}
where $\a$ and $\b$ are spin- and $i$ and $j$ are flavor-indices.
The graded equal-time commutation relations for two $Q$'s and two
$Q^\dagger$'s are analogous.
$\eta_k$ is given by
\begin{equation}
   \eta_k
   = \left\{ 
       \begin{array}{cl}
         1 & \text{for } k=1,2,3 \\
         0 & \text{for } k=4,5,6
       \end{array}
     \right.
.\end{equation}
The quark mass matrix is given by
$m_Q=\diag(m_u,m_d,m_s,m_u,m_d,m_s)$, i.e.,
the fermionic and bosonic quarks have equal masses but different
statistics.  Therefore the contributions of fermionic and
bosonic quarks in virtual quark loops cancel exactly.
The Lagrangian in Eq.~(\ref{eqn:LQQCD}) exhibits a graded 
symmetry 
$[SU(3|3)_L \otimes SU(3|3)_R] \times U(1)_V$
that is assumed to be broken spontaneously to
$SU(3|3)_V \times U(1)_V$.
The dynamics of the emerging 36~pseudo-Goldstone mesons 
can be described at lowest 
order in the chiral expansion by the Lagrangian
\begin{equation}\label{eqn:Lchi}
  {\cal L} =
  \frac{f^2}{8}
    \str\left(\partial^\mu\Sigma^\dagger\partial_\mu\Sigma\right)
    + \l\,\str\left(m_Q\Sigma+m_Q^\dagger\Sigma^\dagger\right)
    + \a\partial^\mu\Phi_0\partial_\mu\Phi_0
    - \mu_0^2\Phi_0^2
\end{equation}
where
\begin{equation} \label{eqn:Sigma}
  \Sigma=\exp\left(\frac{2i\Phi}{f}\right)
  = \xi^2
\end{equation}
and
\begin{equation}
  \Phi=
    \left(
      \begin{array}{cc}
        \pi & \chi^{\dagger} \\ 
        \chi & \tilde{\pi}
      \end{array}
    \right)
.\end{equation}
Here the $\pi$, $\tilde{\pi}$, and $\chi$ are $3\times3$ matrices
of pseudo-Goldstone bosons with quantum numbers of $\bar{q}q$ pairs,
pseudo-Goldstone bosons with quantum numbers of 
$\bar{\tilde{q}}\tilde{q}$ pairs, and pseudo-Goldstone fermions with quantum numbers of $\bar{\tilde{q}}q$ pairs, respectively, 
\begin{equation}
  \pi=
    \left(
      \begin{array}{ccc}
        \eta_u & \pi^+ & K^+ \\ 
        \pi^- & \eta_d & K^0 \\
        K^- & \bar{K^0} & \eta_s
      \end{array}
    \right),\quad
  \tilde{\pi}=
    \left(
      \begin{array}{ccc}
        \tilde{\eta}_u & \tilde{\pi}^+ & \tilde{K}^+ \\ 
        \tilde{\pi}^- & \tilde{\eta}_d & \tilde{K}^0 \\
        \tilde{K}^- & \bar{\tilde{K^0}} & \tilde{\eta}_s
      \end{array}
    \right),
  \quad\text{and}\quad
  \chi=
    \left(
      \begin{array}{ccc}
        \chi_{\eta_u} & \chi_{\pi^+} & \chi_{K^+} \\ 
        \chi_{\pi^-} & \chi_{\eta_d} & \chi_{K^0} \\
        \chi_{K^-} & \chi_{\bar{K^0}} & \chi_{\eta_s}
      \end{array}
    \right)
.\end{equation}
The pion decay constant is fixed by Eq.~(\ref{eqn:Sigma})
and 
$f=132\text{ MeV}$ in QCD.

The flavor-singlet field $\Phi_0$ is defined as  
\begin{equation}
  \Phi_0
  =
  \frac{1}{\sqrt{6}}\str(\Phi)
  =
  \frac{1}{\sqrt{2}}(\eta'-\tilde{\eta}')
\end{equation}
where str() denotes a supertrace over the flavor indices.
$\Phi_0$ is
invariant under $[SU(3|3)_L \otimes SU(3|3)_R] \times U(1)_V$
and thus arbitrary functions of it can be included in the
Lagrangian.
To lowest order in the chiral expansion only the two operators included
in Eq.~(\ref{eqn:Lchi}) with parameters
$\a$ and $\mu_0$ remain and are understood to be inserted 
perturbatively~\cite{Bernard:1992mk}.
Notice that this singlet field $\Phi_0$ is not heavy as in \CPT\
and therefore cannot be 
integrated out. It introduces a new vertex,
the so-called hairpin.

Upon expanding the Lagrangian in Eq.~(\ref{eqn:Lchi}) one finds that
the mesons with quark content $u\bar{u}$, $d\bar{d}$, and $s\bar{s}$,
the only ones relevant for our calculation,
have masses given by
\begin{equation}\label{eqn:mqq}
  m_{qq}^2=\frac{8\lambda m_q}{f^2}
.\end{equation}

\section{Inclusion of Heavy Quarks}
The $D$-mesons with quantum numbers of $c\bar{Q}$ can be written as 
a six-component vector
\begin{equation}
  D=(D_u,D_d,D_s,D_{\tilde{u}},D_{\tilde{d}},D_{\tilde{s}})
.\end{equation} 
Heavy quark symmetry is provided by combining creation and annihilation
operators for the pseudoscalar and vector mesons, 
$D$ and $D^*$ respectively,
together into the field $H^D$ %
\begin{eqnarray}
  H^D&=&\frac{1+\vslash}{2}(\Dslash^*+i\gamma_5D), \\
  \bar{H}^D=\gamma^0H^{D\dagger}\gamma^0
       &=&({\Dslash^*}^\dagger+i\gamma_5D^\dagger)\frac{1+\vslash}{2}
,\end{eqnarray}
where $v$ denotes the velocity of a heavy meson.
In HQET the momentum of a heavy quark is only changed
by a small residual momentum of 
${\mathcal O}(\Lambda_{\text{QCD}})$.
Hence, $v$ is not changed and $H$ is usually denoted by an index
$v$ which we have dropped here to unclutter the formalism.
In the heavy quark limit, 
the dynamics of the heavy mesons are described by the Lagrangian%
~\cite{Booth:1995hx,Sharpe:1996qp}
\begin{equation}\label{eqn:LD}
  {\cal L}_D
  =
  -i\,\tr[\bar{H}_a^Dv_\mu(\partial^\mu\d_{ab}+i V_{ba}^\mu)H_b^D]
  +g\,\tr(\bar{H}_a^DH_b^D\gamma_\nu\gamma_5 A_{ba}^\nu) 
  +\g\,\tr(\bar{H}_a^DH_a^D\gamma_\mu\gamma_5)\,\str A^\mu
\end{equation}
where the traces tr() are over Dirac indices and supertraces str()
over the flavor indices are implicit.
The additional coupling term involving $\Phi_0\sim\str A^\mu$ is a 
feature of \QCPT\ and not present in \CPT.
The light-meson fields are
\begin{equation}
  A_{\mu}
  =
  \frac{i}{2}(\xi^\dagger\partial_\mu\xi-\xi\partial_\mu\xi^\dagger)
  =
  -\frac{1}{f}\partial_\mu\Phi+{\mathcal O}(\Phi^3)
\end{equation}
and
\begin{equation}
  V_{\mu}
  =
  \frac{i}{2}(\xi^\dagger\partial_\mu\xi+\xi\partial_\mu\xi^\dagger)
  =
  \frac{i}{2f^2}[\Phi,\partial_\mu\Phi]+{\mathcal O}(\Phi^4)
.\end{equation}
Expanding the Lagrangian ${\cal L}_D$  to lowest order in the
meson fields leads to the (derivative) 
couplings $DD^*\partial\phi$ and $D^*D^*\partial\phi$
whose coupling constants are equal as a consequence of
heavy quark spin symmetry.
The $DD\partial\phi$ coupling vanishes by parity.

An analogous formalism applies to the fields $B$ and $B^*$ 
which are combined into $H^B$.
Note that the 
axial coupling $g$ is the same for $H^D$ and $H^B$ mesons
as dictated by heavy quark flavor symmetry.

We do not include terms of order $m_q\sim\sqrt{m_\pi}$ 
in the Lagrangian as
explicit chiral symmetry breaking effects are suppressed compared to
the leading corrections.  The presence of these terms is implied by
the nonzero masses $m_{qq}$.

\section{Matrix elements of 
         \protect$\bar{B}^{(*)}\rightarrow D^{(*)}l\bar{\nu}$}
The non-zero hadronic matrix elements for 
$B^{(*)}\rightarrow D^{(*)}$ can be defined 
in terms of the 16 independent form factors 
$h_\pm$, $h_V$, $h_{A_{1,2,3}}$, and $h_{1\dots 10}$ 
as%
~\cite{Falk:1993wt,Manohar:2000dt} 
\begin{equation}\label{eqn:matrixelementfirst}
  \frac{\langle D(v')|\bar{c}\gamma^\mu b|B(v)\rangle}
       {\sqrt{m_Bm_D}}
  =
  h_+(\o)(v+v')^\mu+h_-(\o)(v-v')^\mu 
,\end{equation}
\begin{equation}
  \frac{\langle D^*(v',\e')|\bar{c}\gamma^\mu b|B(v)\rangle}
       {\sqrt{m_Bm_{D^*}}}
  =
  -h_V(\o)\varepsilon^{\mu\nu\a\b}\e'^*_\nu v'_\a v_\b
,\end{equation}
\begin{equation}
  \frac{\langle D^*(v',\e')|\bar{c}\gamma^\mu\gamma_5b|B(v)\rangle}
       {\sqrt{m_Bm_{D^*}}}
  =
  -ih_{A_1}(\o)(\o+1)\e'^{*\mu}
      +ih_{A_2}(\o)(v\cdot\e'^*)v^\mu
      +ih_{A_3}(\o)(v\cdot\e'^*)v'^\mu 
,\end{equation}
\begin{eqnarray}
  \frac{\langle D^*(v',\e')|\bar{c}\gamma^\mu b|B^*(v,\e)\rangle}
       {\sqrt{m_{B^*}m_{D^*}}}
  &=&
  -(\e'^*\cdot \e)[h_1(\o)(v+v')^\mu+h_2(\o)(v-v')^\mu]
  +h_3(\o)(\e'^*\cdot v)\e^\mu
               \nonumber \\
  &&
  +h_4(\o)(\e\cdot v')\e'^{*\mu}
  -(\e\cdot v')(\e'^*\cdot v)[h_5(\o)v^\mu+h_6(\o)v'^\mu]
,\end{eqnarray}
and
\begin{eqnarray}\label{eqn:matrixelementlast}
  \frac{\langle D^*(v',\e')|\bar{c}\gamma^\mu\gamma_5b|B^*(v,\e)\rangle}
       {\sqrt{m_{B^*}m_{D^*}}}
  &=&
  i\varepsilon^{\mu\a\k\d}
  \left\{
    \e^*_\k\e_\d\left[h_7(\o)(v+v')^\mu+h_8(\o)(v-v')^\mu\right]
    \right. \nonumber \\
    &&\left. \phantom{dadasd}
    +v'^\a v^\b
     \left[h_9(\o)(\e'^*\cdot v)\e^\mu+h_{10}(\o)(\e\cdot v')\e'^{*\mu}\right]
  \right\}
.\end{eqnarray}
Here, $\o=v\cdot v'$ and
$v$ [$\e$] and $v'$ [$\e'$]
are the velocities [polarization vectors]
of the initial state $B^{(*)}$ meson
and final state $D^{(*)}$ meson, respectively.
Note that we will not explicitly calculate matrix elements
of $B^*\to D$ as these can be easily related to the
$B\to D^*$ calculation by a hermitian conjugation of the matrix
elements and an interchange of the $c$ and $b$ quarks,
i.e., $B^{(*)}\leftrightarrow D^{(*)}$.

In the heavy quark limit the matrix elements
in Eqs.\ (\ref{eqn:matrixelementfirst})--(\ref{eqn:matrixelementlast})
are reproduced by the operator
\begin{equation}
  \bar{c}\gamma^\mu(1-\gamma_5)b
  \rightarrow 
  -\xi(\omega)\tr[\bar{H}^D_{v'}\gamma^\mu(1-\gamma_5)H^B_v]
.\end{equation}
Here, $\xi(\o)$
is the universal Isgur-Wise function~\cite{Isgur:1989vq,Isgur:1990ed}
with the normalization $\xi(1)=1$.
To lowest order in the heavy quark expansion one finds
\begin{equation}
  h_+(\o)
  =h_V(\o)
  =h_{A_1}(\o)=h_{A_3}(\o)
  =h_1(\o)=h_3(\o)=h_4(\o)=h_7(\o)=\xi(\o)
\end{equation}
and the remaining 8 form factors vanish.

The discussion of the $\bar{B}^{(*)}\rightarrow D^{(*)}l\bar{\nu}$
matrix elements is similar for different flavors of the light quark $q$
content of the $B^{(*)}$ and $D^{(*)}$ mesons;
it applies equally to $q=u$, $d$, or~$s$
as the theory splits into three similar copies
of a one-flavor theory.
In the limit of light quark $SU(3)_V$ flavor symmetry the matrix elements
(and in particular the Isgur-Wise function) are therefore
independent of the
light quark flavor.
However, in nature the masses of the $u$, $d$, and $s$ quarks are
different and $SU(3)_V$ is not an exact symmetry.
Therefore our results will include terms that 
depend upon $m_q$ via the meson masses
$m_{qq}$ defined in Eq.~(\ref{eqn:mqq}).

\section{\protect$1/M$ Corrections}
The lowest order heavy quark symmetry violating operator 
that can be included in the Lagrangian ${\mathcal L}_D$ in
Eq.~(\ref{eqn:LD}) is the dimension-three operator 
$\frac{\l_{D_2}}{M_D}
  \tr\left[\bar{H_a}^D\sigma^{\mu\nu}H_a^D\sigma_{\mu\nu}\right]$.
It violates heavy-quark spin and flavor symmetries and
comes from the QCD magnetic moment operator
$\bar{c}\sigma^{\mu\nu}G_{\mu\nu}^AT^Ac$,
where $G_{\mu\nu}^A$ is the gluon field strength tensor and
$T^A$ with $A=1\dots8$ are the eight color $SU(3)$ generators.
This operator gives rise to a
mass difference between the $D$ and $D^*$
mesons of
\begin{equation}
  \D_D=m_{D^*}-m_D=-8\frac{\l_{D_2}}{\bar{M}_D}
.\end{equation}
This effect can be taken into account by modifying the $D$ and $D^*$
propagators which become
\begin{equation}
  \frac{i\d_{ab}}{2(v\cdot k+3\D_D/4+i\e)}
  \quad\text{and}\quad
  \frac{-i\d_{ab}(g_{\mu\nu}-v_\mu v_\nu)}{2(v\cdot k-\D_D/4+i\e)}
,\end{equation}
respectively,
so that in the rest frame, where $v=(1,0,0,0)$, an on-shell $D$ has
residual energy of $-3\D_D/4$ and an on-shell $D^*$ has residual
energy of $\D_D/4$.
A similar effect due to the inclusion of
a QCD magnetic moment operator for the $b$ quark
applies to the $B^{(*)}$ mesons.

There are no corrections to the matrix elements for the semileptonic decays
$B^{(*)}\rightarrow D^{(*)}e\nu$ of $\order(1/M)$ at zero-recoil 
according to Luke's theorem~\cite{Luke:1990eg}.  
The leading corrections enter at $\order(1/M^2)$.
In addition to tree-level contributions from the insertion of 
$\order(1/M^2)$ suppressed operators 
into the heavy quark Lagrangian or the current 
there 
are one-loop contributions from wavefunction renormalization
and vertex correction.
These one-loop diagrams have a non-analytic dependence on the
meson mass $m_{qq}$ and depend on the subtraction point $\mu$.
This dependence on $\mu$ is canceled by the tree-level contribution
of the $\order(1/M^2)$ operators.

Because of the absence of disconnected
quark loops in QQCD, which manifests itself
as a cancellation between intermediate pseudo-Goldstone bosons and 
pseudo-Goldstone fermions in loops in \QCPT, the only loop diagrams
that survive are those that contain a hairpin interaction or 
a $\g$ coupling.

\begin{figure}[tb]
    \includegraphics[width=\textwidth]{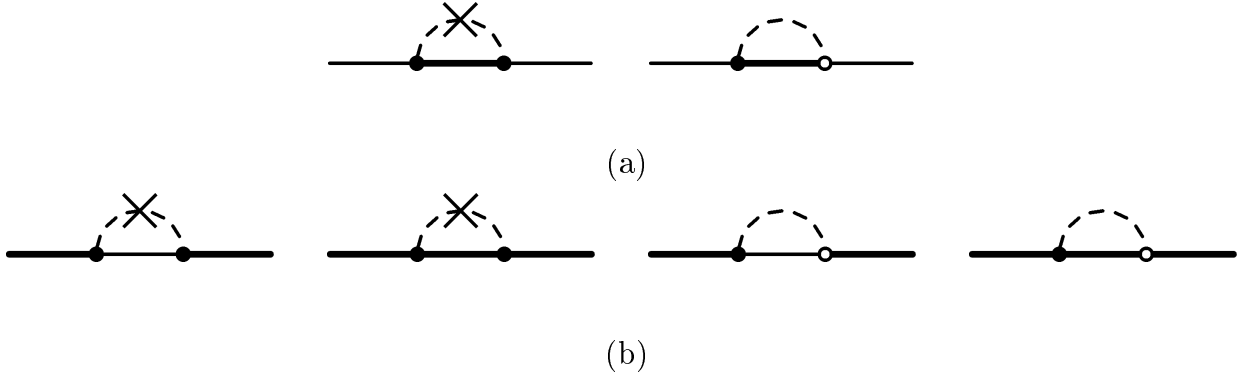}%
    \caption{
      Graphs contributing to wavefunction renormalization for heavy
      (a) pseudoscalar and (b) vector mesons.
      A thin [thick] line denotes a heavy pseudoscalar [vector] meson,
      a dashed line denotes the $\Phi_0$, while a dashed-crossed
      line denotes the insertion of a hairpin.
      A full [empty] vertex denotes a $g$ [$\g$]
      coupling.
  }
  \label{F:wf-renorm}
.\end{figure}

The wavefunction renormalization contributions for
the pseudoscalar and vector meson, $Z_{D/B}$ and $Z^*_{D/B}$, 
respectively, 
come from the one-loop diagrams shown in 
Fig.~(\ref{F:wf-renorm}) and
have been calculated in%
~\cite{Booth:1994rr,Booth:1995hx,Sharpe:1996qp}.
Including the $\a$ coupling we find for these diagrams
\begin{eqnarray}
  Z
  &=&
    1
    +\frac{ig^2\mu_0^2}{f^2}H_1(\D)
    -\frac{ig^2\a}{f^2}H_2(\D)
    +\frac{6i\g g}{f^2}F_1(\D)
\end{eqnarray}
and
\begin{eqnarray}
  Z^*
  &=&
    1
    +\frac{ig^2\mu_0^2}{3f^2}H_1(-\D)
    -\frac{ig^2\a}{3f^2}H_2(-\D)
    +\frac{2i\g g}{f^2}F_1(-\D)
   \nonumber \\
  &&   
    +\frac{2ig^2\mu_0^2}{3f^2}H_1(0)
    -\frac{2ig^2\a}{3f^2}H_2(0)
    +\frac{4i\g g}{f^2}F_1(0)
.\end{eqnarray}
The functions $H_1$, $H_2$, and $F_1$ come from loop integrals and
are given in the Appendix.
Note that in the heavy quark limit where $\D=0$
one recovers
$Z=Z^*$, as required by heavy quark symmetry.

\begin{figure}[tb]
    \includegraphics[width=\textwidth]{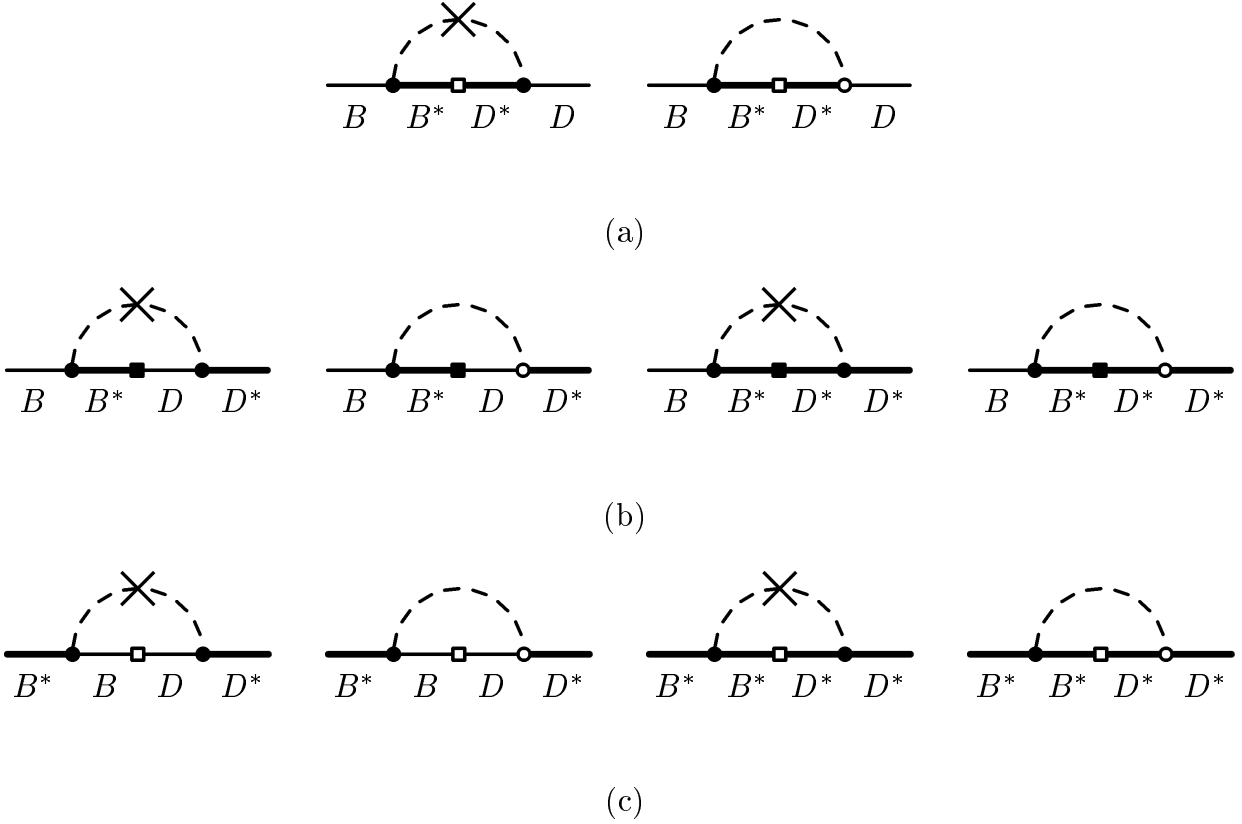}%
    \caption{
      \QCPT\ graphs which contribute to the vertex correction 
      of the form factors 
      (a) $h_+(1)$, (b) $h_{A_1}(1)$, and (c) $h_1(1)$.
      A full [empty] square denotes the insertion of the
      operator $\bar{c}\g^\mu\g_5b$ [$\bar{c}\g^\mu b$]. 
  }
  \label{F:vertex-corr}
\end{figure}

The vertex corrections come from one-loop diagrams.
The non-vanishing contributions
are shown in Fig.~(\ref{F:vertex-corr}).
Combining the wavefunction renormalization and vertex corrections 
and including a local counterterm to cancel the dependence
on the renormalization scale $\mu$ we 
find the following corrections for the form factors:
\begin{eqnarray}\label{eqn:hplus}
  \d h_+(1)
  &=&
  X_+(\mu)
  +\frac{Z_B-1}{2}+\frac{Z_D-1}{2}
               \nonumber \\
              &&
    -\frac{ig^2}{f^2}
    \left[
      \mu_0^2\,H_5(\D_B,\D_D)-\a\,H_8(\D_B,\D_D)
    \right]
    -\frac{6ig\g}{f^2}G_5(\D_B,\D_D)
  \nonumber \\
  &\rightarrow&
  X_+(\mu)
  +\frac{1}{(4\pi f)^2}
  \left(
    \frac{g^2\mu_0^2}{3m^2}
   -\left[
      \frac{g^2\a}{3}-2g\g
    \right]
    \log\frac{m^2}{\mu^2}
  \right)
  (\D_B-\D_D)^2
               \nonumber \\
              &&
  +{\mathcal O}(\{\D_B,\D_D\}^3)
,\end{eqnarray}
\begin{eqnarray}\label{eqn:hA1}
  \d h_{A_1}(1)
  &=&
  X_{A_1}(\mu)
  +\frac{Z_B-1}{2}+\frac{Z_D^*-1}{2}
               \nonumber \\
              &&
    -\frac{ig^2}{3f^2}
    \left[
      \mu_0^2\,H_5(\D_B,-\D_D)-\a\,H_8(\D_B,-\D_D)
    \right]
    -\frac{2ig\g}{f^2}G_5(\D_B,-\D_D)
               \nonumber \\
              &&
    -\frac{2ig^2}{3f^2}
    \left[
      \mu_0^2\,H_5(\D_B,0)-\a\,H_8(\D_B,0)
    \right]
    -\frac{4ig\g}{f^2}G_5(\D_B,0)
  \nonumber \\
  &\rightarrow&
  X_{A_1}(\mu)
   +\frac{1}{(4\pi f)^2}
   \left(
     \frac{g^2\mu_0^2}{9m^2}
    -\left[
       \frac{g^2\a}{9}-\frac{2g\g}{3}
     \right]
     \log\frac{m^2}{\mu^2}
   \right)
   \left(3\D_B^2+\D_D^2+2\D_B\D_D\right)
               \nonumber \\
              &&
  +{\mathcal O}(\{\D_B,\D_D\}^3)
,\end{eqnarray}
and
\begin{eqnarray}\label{eqn:h1}
  \d h_1(1)
  &=&
  X_1(\mu)
  +\frac{Z_B^*-1}{2}+\frac{Z_D^*-1}{2}
               \nonumber \\
              &&
    -\frac{ig^2}{3f^2}
    \left[
      \mu_0^2\,H_5(-\D_B,-\D_D)-\a\,H_8(-\D_B,-\D_D)
    \right]
    -\frac{2ig\g}{f^2}G_5(-\D_B,-\D_D)
               \nonumber \\
              &&
    -\frac{2ig^2}{3f^2}
    \left[
      \mu_0^2\,H_5(0,0)-\a\,H_8(0,0)
    \right]
    -\frac{4ig\g}{f^2}G_5(0,0)
  \nonumber \\
  &\rightarrow&
  X_1(\mu)
   +\frac{1}{(4\pi f)^2}
   \left(
     \frac{g^2\mu_0^2}{9m^2}
    -\left[
       \frac{g^2\a}{9}-\frac{2g\g}{3}
     \right]
     \log\frac{m^2}{\mu^2}
   \right)
   \left(\D_B-\D_D\right)^2
               \nonumber \\
              &&
  +{\mathcal O}(\{\D_B,\D_D\}^3)
,\end{eqnarray}
which are defined by $h_+(1)=1+\d h_+(1)$
and analog expressions for $\d h_{A_1}(1)$ and $\d h_1(1)$.
The functions $H_5$, $H_8$, and $G_5$ come from loop-integrals 
that are 
listed in the Appendix and we have defined $m=m_{qq}$.
The insertions of tree-level $\order(1/M^2)$ operators
are represented by 
the functions $X_+(\mu)$, $X_{A_1}(\mu)$, and $X_1(\mu)$
which are independent of $m$  and exactly cancel the
$\mu$ dependence of the logarithm.
These functions can be extracted from lattice simulations by measuring
the zero-recoil form factors for a varying mass of the light quark.

Experimentally, 
$\D_D\approx 142\,\text{MeV}$ and 
$\D_B\approx 46\,\text{MeV}$ so that the ratios
$\D_D/m$ and $\D_B/m$,
which enter the form factor corrections through
the function
$R(\D/m)$ (defined in the Appendix), 
are $\order(1)$.
On the lattice, however, one can vary all 
quark masses.
Expanding first in powers of $\D$ and then taking the chiral limit
$m\to 0$
one finds the formal limits given in 
Eqs.\ (\ref{eqn:hplus})--(\ref{eqn:h1})
where we have only kept the pieces non-analytic in $m$.
This demonstrates 
that the terms linear in $\D_D$ and $\D_B$, although
present in wavefunction renormalization and vertex corrections,
cancel as required by Luke's theorem~\cite{Luke:1990eg}.
The leading order corrections are 
${\mathcal O}(\{\D_B,\D_D\}^2)$.

As a consistency check one can 
restore heavy quark flavor symmetry by taking $\D_B=\D_D$.
Since the ${\mathcal O}(1/M^2)$ corrections to $h_+(1)$ and $h_1(1)$
are proportional to $(\D_B-\D_D)^2$ they disappear as they should
since the charge associated with the operators
$\bar{c}\g_\mu c$ and $\bar{b}\g_\mu b$
is conserved.
This argument does not apply for the $B\to D^*$ transition matrix 
element in the limit $\D_B=\D_D$ since there is no conserved axial charge
associated with the operators 
$\bar{c}\g_\mu\g_5c$ and $\bar{b}\g_\mu\g_5b$.

In the chiral limit, 
the term proportional
to $\mu_0^2$
has a $1/m^2$ singularity and
dominates over the terms proportional
to $\a$ and $\g$ that are only log-divergent.  
This is analogous to a term of the form
$(m_{qq}^2-m_{jj}^2)/m_{qq}^2$ found by 
Savage~\cite{Savage:2001jw}
for \PQCPT\ (here, $m_{qq}$ and $m_{jj}$ are valence and sea quark 
masses, respectively).
In the limit $m_{jj}\to m_{qq}$ this term, however, vanishes as 
\PQCPT\ goes to \CPT\ where the dominant term is $\log m_{qq}$. 
In \QCPT, on the other hand, the
$1/m_{qq}^2$ pole
persists, revealing the sickness
of QQCD where the hairpin interactions give a completely
different chiral behavior than in QCD.

The size of $\mu_0$ can be estimated from the 
$\eta$-$\eta'$ mass splitting~\cite{Sharpe:1992ft}, 
large $N_C$ arguments~\cite{Witten:1979vv,Veneziano:1979ec}
($N_C$ being the number of colors),
or lattice calculations.
These estimates imply 
$\mu_0\approx 500-900\,\text{MeV}$;
for the purpose of dimensional analysis we use
$\mu_0\sim \order(\LQCD)$.
Taking $g\sim \order(1)$
we therefore find that 
$\d h_+$, $\d h_{A_1}$, and $\d h_1$ are of order
$\D^n/m^n\sim \LQCD^{3n/2}/(M^nm_q^{n/2})$ for $n\ge 2$
and thus larger than tree-level heavy quark symmetry breaking operators
that are suppressed by $\LQCD/M$.

To show the dependence of the zero-recoil form-factors 
on the 
mass of the light spectator quark it is necessary to know the 
numerical values of the parameters $\mu_0$, $g$, $\a$, and $\g$.
In determining reasonable values for these couplings we 
follow the discussion by Sharpe and Zhang~\cite{Sharpe:1996qp}.
Assuming that $g$ is similar to the \CPT\ value we use
$g^2=0.4$.
The hairpin coupling $\a$ is proportional to $1/N_C$, 
and thus assumed to be small; we use two values, $\a=0$ and $\a=0.7$.
The coupling $\g$ is known to be suppressed by
$1/N_C$ compared to $g$, the sign is undetermined. 
We take $-g\le\g\le g$
(see~\cite{Sharpe:1996qp} and references therein).

\begin{figure}[tb]
    \includegraphics[width=0.8\textwidth]{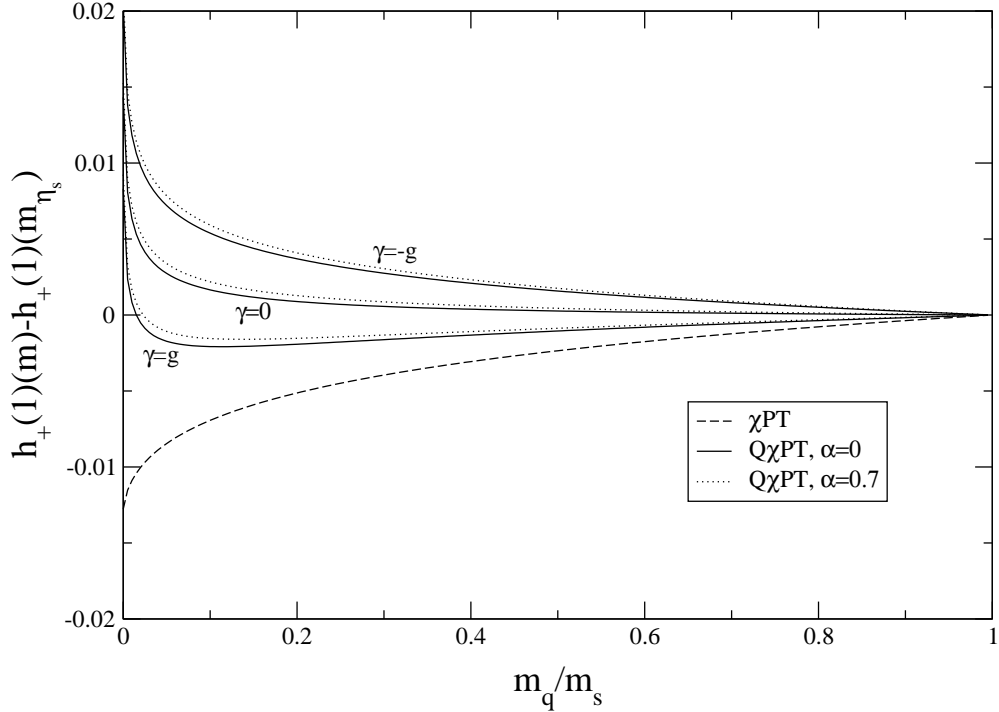}%
    \caption{
      Dependence of $h_+(1)$ on the mass $m_q$ of the light spectator quark 
      in \QCPT.  For comparison, the \CPT\ result from%
      ~\cite{Randall:1993qg} is also shown (dashed line).
      The result has been normalized to unity for $m_q=m_s$.
      We have chosen $\mu_0=700\,\text{MeV}$ and
      $g^2=0.4$.}
  \label{F:plus-plot}
\end{figure}

\begin{figure}[tb]
    \includegraphics[width=0.8\textwidth]{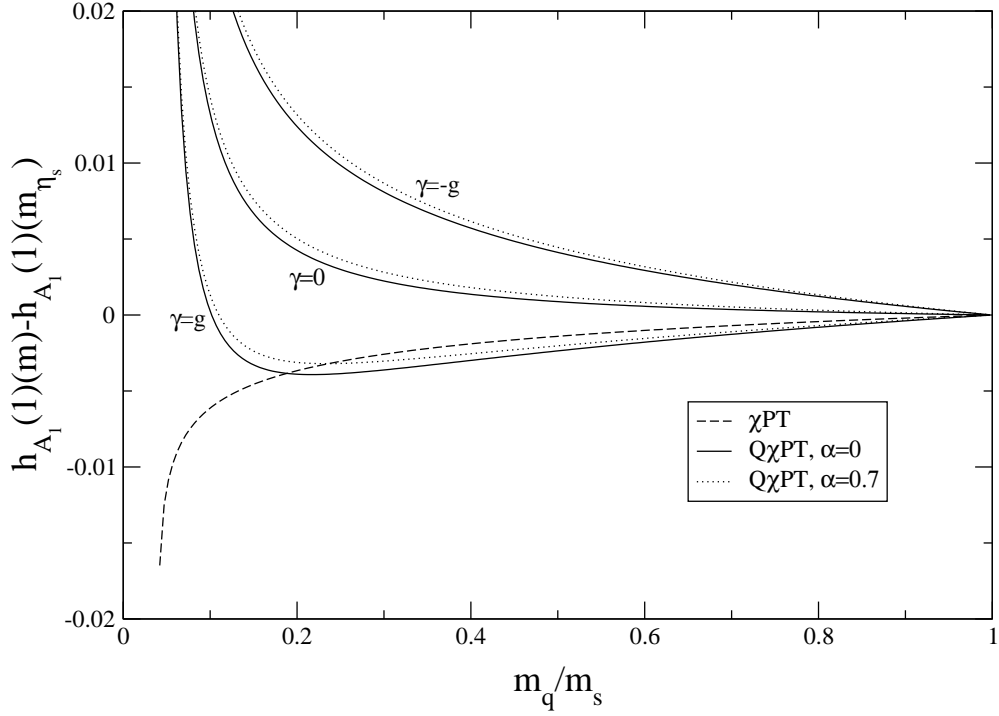}%
    \caption{
      Dependence of $h_{A_1}(1)$ on the mass of the light spectator quark
      in \QCPT.  The dashed line denotes the \CPT\ result%
      ~\cite{Randall:1993qg}.
      The numerical values for the parameters are those used in
      Fig.~(\ref{F:plus-plot}).}
  \label{F:A1-plot}
\end{figure}

With these parameters, 
the dependence of $h_+(1)$ and $h_{A_1}(1)$ on the 
mass of the light spectator quark $m_q$
is show in 
Figs.\ (\ref{F:plus-plot}) and (\ref{F:A1-plot}),
respectively.
The graphs are plotted against $m_q$ 
in units of the 
strange quark mass $m_s$ with $m_q/m_s=m^2/m_{\eta_s}^2$ where 
$m_{\eta_s}^2=2m_K^2$.
The behavior of $h_+(1)$ in \QCPT\
is dominated at small $m$ by the $1/m^2$ pole 
that is non-existent in \CPT.
Lattice calculations of $h_+(1)$%
~\cite{El-Khadra:2001rv} 
show
a small downward trend for 
decreasing $m_q$ down to the chiral limit that is 
similar to the downward trend 
seen from the \CPT\ calculation (dashed line).
The same behavior (down to $m_q\approx0.1 m_s$)
can also be seen for \QCPT\ for a certain choice of
parameters (e.g., $\g$ positiv).
The case of $h_{A_1}(1)$ is different as there is a pole at $m=\D_D$ 
which is close to the physical pion mass.  Here, 
both $D^*$ and $\pi$ can be on-shell and 
the decay $B\to D^*\pi$ becomes kinematically allowed.
Lattice calculations of $h_{A_1}(1)$%
~\cite{Hashimoto:2001nb} for $m_q=(0.6\dots1) m_s$
show
a small downward trend for  
decreasing $m_q$ similar to the downward trend 
seen from the \CPT\ calculation [dashed line in Fig.~(\ref{F:A1-plot})].
A similar trend down to $m_q\approx0.2 m_s$
can also be seen in the \QCPT\ calculation 
for a relatively large positiv value 
of $\g$.

Although the downward trend in the lattice data for the two cases 
seems significant as 
the statistical errors are highly correlated, the 
uncertainty is still relatively high (typically $\pm0.01$)
and the existing lattice data can be accommodated by a wide
range of values for the parameters in the \QCPT\ Lagrangian.

Finally, we calculate the double ratios defined in
Eqs. (\ref{eqn:Rplus})--(\ref{eqn:RA1}) using the 
results in Eqs.\ (\ref{eqn:hplus})--(\ref{eqn:h1}).
We find 
\begin{equation}
  {\mathcal R}_+
  =
  1+2\d h_+(1)
,\end{equation}
\begin{equation}
  {\mathcal R}_1
  =
  1+2\d h_1(1)
,\end{equation}
and
\begin{eqnarray}
  {\mathcal R}_{A_1}
  &=&
  1
  +
  \tilde{X}_{A_1}(\mu)
               \nonumber \\
               &&
  -\frac{ig^2}{3f^2}
   \left\{
     \mu_0^2
     \left[
       H_5(\D_B,-\D_D)+H_5(\D_D,-\D_B)-H_5(\D_D,-\D_D)-H_5(\D_B,-\D_B)
     \right]
               \right.\nonumber \\
               && \left.
     \phantom{gggggg}
     -\a
     \left[
       H_8(\D_B,-\D_D)+H_8(\D_D,-\D_B)-H_8(\D_D,-\D_D)-H_8(\D_B,-\D_B)
     \right]     
   \right\}
               \nonumber \\
               &&
   -\frac{2ig\g}{f^2}
     \left[
       G_5(\D_B,-\D_D)+G_5(\D_D,-\D_B)-G_5(\D_D,-\D_D)-G_5(\D_B,-\D_B)
     \right]    
  \nonumber \\
  &\rightarrow&
  1
  +
  \tilde{X}_{A_1}(\mu)
   -\frac{1}{(4\pi f)^2}
   \left(
     \frac{2g^2\mu_0^2}{9m^2}
    -\left[
       \frac{2g^2\a}{9}-\frac{4g\g}{3}
     \right]
     \log\frac{m^2}{\mu^2}
   \right)
   \left(\D_B-\D_D\right)^2
               \nonumber \\
              &&
  +{\mathcal O}(\{\D_B,\D_D\}^3)
,\end{eqnarray}
where 
$\tilde{X}_{A_1}(\mu)$
is the counter term associated with ${\mathcal R}_{A_1}$.

\section{Conclusions}
Knowledge of the $B^{(*)}\to D^{(*)}$ form factors
at the zero-recoil point
is crucial to extract the value of $V_{cb}$ from experiment.
In the limit that the heavy quarks are infinitely heavy
HQET predicts that the form factors 
$h_+$, $h_{A_1}$, and $h_1$ are equal, $h_+(1)=h_{A_1}(1)=h_1(1)=\xi(1)$.
The formally dominant 
correction due to breaking of heavy quark symmetry
comes from the inclusion of a $\order(1/M)$ dimension-three
operator in the Lagrangian that leads 
to
hyperfine-splitting between the heavy pseudoscalar and vector mesons. 
These leading order corrections are ${\mathcal O}(\{\D_B,\D_D\}^2$
as required by Luke's theorem.

Recent lattice simulations using the quenched approximation of
QCD have made a big step forward in determining these zero-recoil
form factors.  Presently, however, the simulations use
light quark masses that are much heavier than the physical ones
and therefore rely on a chiral extrapolation down
to the physical quark masses. 
In this paper we have calculated the dominant corrections
to the form factors $h_+$, $h_{A_1}$, and $h_1$
in \QCPT\ and determined the non-analytic dependence 
on the light quark masses via the light meson masses $m_{qq}$. 
Using these results, instead of the \CPT\ calculation, 
to extrapolate the
QQCD lattice measurements of these form factors
down to the physical pion mass should give a more reliable
estimate of the errors associated with the chiral extrapolation.

We have also calculated the corrections to certain double ratios
that are used in lattice QCD calculations of the decay
$B\to D^*$.

\acknowledgments
I would like to thank Martin Savage for many
helpful discussions.
I am also grateful to him, Andreas Kronfeld, and 
Steve Sharpe for useful comments on the manuscript.
This work is supported in part by the U.S.\ Department of Energy
under Grant No.\ DE-FG03-97ER4014.

\appendix
\section{Integrals}
We list the functions $H_1$, $H_2$, $F_1$, $H_5$, $H_8$, and $G_5$
even though
some of them have appeared in the literature before%
~\cite{Boyd:1995pa,Boyd:1995pq}.
Here, $m=m_{qq}$ is the mass of the $q\bar{q}$ light meson 
in the loop
where $q=u$, $d$, or~$s$ is the light (spectator) quark content
of the heavy mesons.
We have used dimensional regularization 
with the $\overline{\text{MS}}$ scheme,
where $1/\e'\equiv 1/\e-\g_E+\log 4\pi+1$. At the end we 
set $1/\e'=0$.
As a shorthand we have defined the function
\begin{equation}
  R(x)
  =
  \sqrt{x^2-1}
  \log\left(\frac{x-\sqrt{x^2-1+i\e}}{x+\sqrt{x^2-1+i\e}}\right),
\end{equation}
which occurs frequently.
We also need its derivative $dR/dx$ given by
\begin{equation}
  R'(x)=\frac{x}{x^2-1}R(x)-2
.\end{equation}

For the calculation of the wavefunction renormalization contribution
we need the derivatives of the loop integrals for the
diagrams in Fig.~(\ref{F:wf-renorm}):
\begin{equation}
  H_1(\D)
  =
  \frac{i}{16\pi^2}
  \left[
    \log\frac{m^2}{\mu^2}-\frac{1}{\e'}-1
    -R'\left(\frac{\D}{m}\right)
  \right]
,\end{equation}
\begin{eqnarray}
  H_2(\D)
  &=&
  \frac{i}{16\pi^2}
  \left[
    \frac{16}{3}\D^2-\frac{10}{3}m^2
    +2(m^2-\D^2)\left(\!\log\frac{m^2}{\mu^2}-\frac{1}{\e'}\right)
    +\frac{4}{3}\D mR\left(\frac{\D}{m}\right)
  \right.     \nonumber \\
    && \left. \phantom{dddddd}
    +\left(\frac{2}{3}\D^2-\frac{5}{3}m^2\right)R'\left(\frac{\D}{m}\right)
  \right]
,\end{eqnarray}
and
\begin{eqnarray}
  F_1(\D)
  &=&
  \frac{i}{16\pi^2}
  \left[
    \frac{10}{3}\D^2-\frac{4}{3}m^2
    +\left(m^2-2\D^2\right)\left(\!\log\frac{m^2}{\mu^2}-\frac{1}{\e'}\right)
    +\frac{4}{3}\D mR\left(\frac{\D}{m}\right)
               \right .\nonumber \\
              && \left. \phantom{dddddd}
    +\frac{2}{3}(\D^2-m^2)R'\left(\frac{\D}{m}\right)
  \right]
.\end{eqnarray}

For the loop-integrals of the vertex corrections
one finds
\begin{equation}
  H_5(\D,\Dtilde)
  = 
  \frac{i}{16\pi^2}
  \left(
    \log\frac{m^2}{\mu^2}-\frac{1}{\e'}
    -1
    -\frac{m}{\D-\Dtilde}
     \left[R\left(\frac{\D}{m}\right)-R\left(\frac{\Dtilde}{m}\right)\right]
  \right) 
,\end{equation}
\begin{eqnarray}
  H_8(\D,\Dtilde)
  &=&
  \frac{i}{16\pi^2}
  \left(  
    \left[2m^2-\frac{2}{3}(\D^2+\D\Dtilde+\Dtilde^2)\right]
     \left(\log\frac{m^2}{\mu^2}-\frac{1}{\e'}\right)   
    +\frac{16}{9}(\D^2+\D\Dtilde+\Dtilde^2)
  \right.     \nonumber \\
    && \left. \phantom{dddddd}
    -\frac{10}{3}m^2
    +\frac{m(5m^2-2\Dtilde^2)}{3(\D-\Dtilde)}R\left(\frac{\Dtilde}{m}\right)
    -\frac{m(5m^2-2\D^2)}{3(\D-\Dtilde)}R\left(\frac{\D}{m}\right)
  \right) 
,\end{eqnarray}
and
\begin{eqnarray}
  G_5(\D,\Dtilde)
  &=&
  \frac{i}{16\pi^2}
  \left(
    \frac{10}{9}(\D^2+\D\Dtilde+\Dtilde^2)-\frac{4}{3}m^2     
    +\left[m^2-\frac{2}{3}(\D^2+\D\Dtilde+\Dtilde^2)\right]
     \left(\!\log\frac{m^2}{\mu^2}-\frac{1}{\e'}\right)
  \right.     \nonumber \\
    && \left. \phantom{dddddd}
    +\frac{2m(\D^2-m^2)}{3(\D-\Dtilde)}R\left(\frac{\D}{m}\right)
    -\frac{2m(\Dtilde^2-m^2)}{3(\D-\Dtilde)}R\left(\frac{\Dtilde}{m}\right)
  \right) 
.\end{eqnarray}


\end{document}